\documentclass[aps,pra,twocolumn]{revtex4-1}%revtex4-1
\usepackage{amsmath}
\usepackage{amsfonts}
\usepackage{amssymb}
\usepackage{graphicx}
\usepackage[colorlinks,linkcolor=blue,anchorcolor=blue,citecolor=blue,urlcolor=blue]{hyperref}
\usepackage{mathrsfs}
\usepackage{epsfig}
\usepackage{float}
\usepackage{subfigure}
\usepackage{mathtools}

\begin{document}

\title{Investigation on the Cs \textbf{$6S_{1/2}$} to \textbf{$7D$} electric quadrupole transition via monochromatic two-photon process at 767 nm}
\author{Sandan Wang$^{1,2}$}
\author{Jinpeng Yuan$^{1,2}$}
\email{yjp@sxu.edu.cn}
\author{Lirong Wang$^{1,2}$}
\email{wlr@sxu.edu.cn}
\author{Liantuan Xiao$^{1,2}$}
\author{Suotang Jia$^{1,2}$}

\affiliation{$^{1}$State Key Laboratory of Quantum Optics and Quantum Optics Devices, Institute of Laser Spectroscopy, Shanxi University, Taiyuan 030006, People's Republic of China}
\affiliation{$^{2}$Collaborative Innovation Center of Extreme Optics, Shanxi University, Taiyuan 030006, People's Republic of China}

\begin{abstract}
We experimentally demonstrate the cesium electric quadrupole transition from the $6S_{1/2}$ ground state to the $7D_{3/2, 5/2}$ excited state through a virtual level by using a single laser at 767 nm. The excited state energy level population is characterized by varying the laser power, the temperature of the vapor, and the polarization combinations of the laser beams. The optimized experimental parameters are obtained for a high resolution transition interval identification. The magnetic dipole coupling constant $A$ and electric quadrupole coupling constant $B$ for the $7D_{3/2, 5/2}$ states are precisely determined by using the hyperfine levels intervals. The results, $A$ = 7.39 (0.06) MHz, $B$ = -0.19 (0.18) MHz for the $7D_{3/2}$ state, and $A$ = -1.79 (0.05) MHz, $B$ =1.05 (0.29) MHz for the $7D_{5/2}$ state, are in good agreement with the previous reported results. This work is beneficial for the determination of atomic structure information and parity non-conservation, which paves the way for the field of precision measurements and atomic physics.
\end{abstract}

\maketitle
\section{Introduction}

The precise measurement of atomic structure properties has laid a good experimental foundation for the development of atomic physics and quantum mechanics \cite{Wood1997,Porsev2009,Zheng2017,Yuan2019,Li2020}. The hyperfine splitting (HFS), resulting from electron-nuclear interactions, can be used to determine the strength of the interaction of magnetic dipoles, electric quadruple, and magnetic octuplet interactions between the nucleus and the orbital electrons \cite{Ramos2019}. Further, the measurement of the excited states hyperfine splitting can help in the construction and modification of the atomic wave functions \cite{Arimondo1977}, which is beneficial for understanding the atomic structure and fundamental physics. The hyperfine coupling constants measurement is achieved by using the high precision spectroscopy technology, which has various presentations such as saturated absorption spectroscopy \cite{Udem2000}, polarization spectroscopy \cite{Kirkbride2014}, electromagnetically induced transparency \cite{J. Yuan2019}, and two-photon transition spectroscopy \cite{Ray2020}.

Recently, the $6S-7D$ stepwise excitation transition spectroscopy of the cesium atom is widely researched with the motivation of high precision frequency standard. The $6S-7D$ two-photon transition has a narrow natural linewidth because the $7D$ state energy level has a long lifetime of about 160 ns \cite{Safronova2016}. Meanwhile, the time interval that the radio frequency oscillates 9192631770 times, corresponding to the transition frequency interval between the two-ground state hyperfine energy levels of the $^{133}$Cs atoms, is defined to be 1 second. Therefore, the Cs $6S-7D$ two-photon transition high-precision spectroscopy has the potential to serve as a secondary frequency standard for a frequency-doubled 1534 nm laser in the C-band window of quantum telecommunication \cite{Heshamia2016}.

Several experimental platforms are demonstrated for the Cs $6S-7D$ two-photon transition spectroscopy by two laser beams coupling with the real energy levels. The electric field induced hyperfine level-crossings of $6S_{1/2}-7D$ two-step laser excitation are illustrated and the hyperfine constants of the $7D_{5/2}$ state are obtained in cesium \cite{Auzinsh2006,Auzinsh2007}. The $7D_{3/2}$ hyperfine coupling constants are measured using two-photon fluorescence spectroscopy by a photomultiplier tube \cite{Kortyna2008}. The direct frequency comb spectroscopy is performed in cesium vapor with a high precision mode-locked laser \cite{Stalnaker2010}. However, the above-mentioned systems are complicated and expensive. A simple and effective mechanism is proposed with obvious advantages for the potential frequency standard. The $6S-7D$ monochromatic two-photon transition has some advantages, such as superior Doppler-free background caused by counter-propagation geometry, more candidates for frequency standards due to the transition selection rule $|\bigtriangleup$$F|$$\leq$2, and more precise determination of hyperfine coupling constant via more energy level intervals identification. Therefore, Doppler-free two-photon transitions of cesium $6S_{1/2}-7D_{3/2, 5/2}$ are observed in a thermal vapor \cite{Lee2011}. And the measurement of the hyperfine structure in the $7D_{3/2}$ state is obtained by one-color two-photon fluorescence spectroscopy \cite{Kumar2013}. Further deeper and detailed exploration is required for the monochromatic two-photon transition process in the field of precision measurement.

In this paper, we report on the study of the $6S_{1/2}-7D_{3/2, 5/2}$ electric quadrupole transition in a thermal Cs vapor by using a single laser with 767 nm. We explore the dependency of the atomic hyperfine structure spectral profile on the vapor temperature, the laser power, and the polarization combinations of the laser beams. Thus, the hyperfine coupling constants of $7D_{3/2, 5/2}$ states can be obtained via hyperfine structure level intervals. This research can provide a solid experimental basis for the potential application of precision measurement and quantum telecommunications.

\begin{figure}[ht!]
	\includegraphics[width=3.5in]{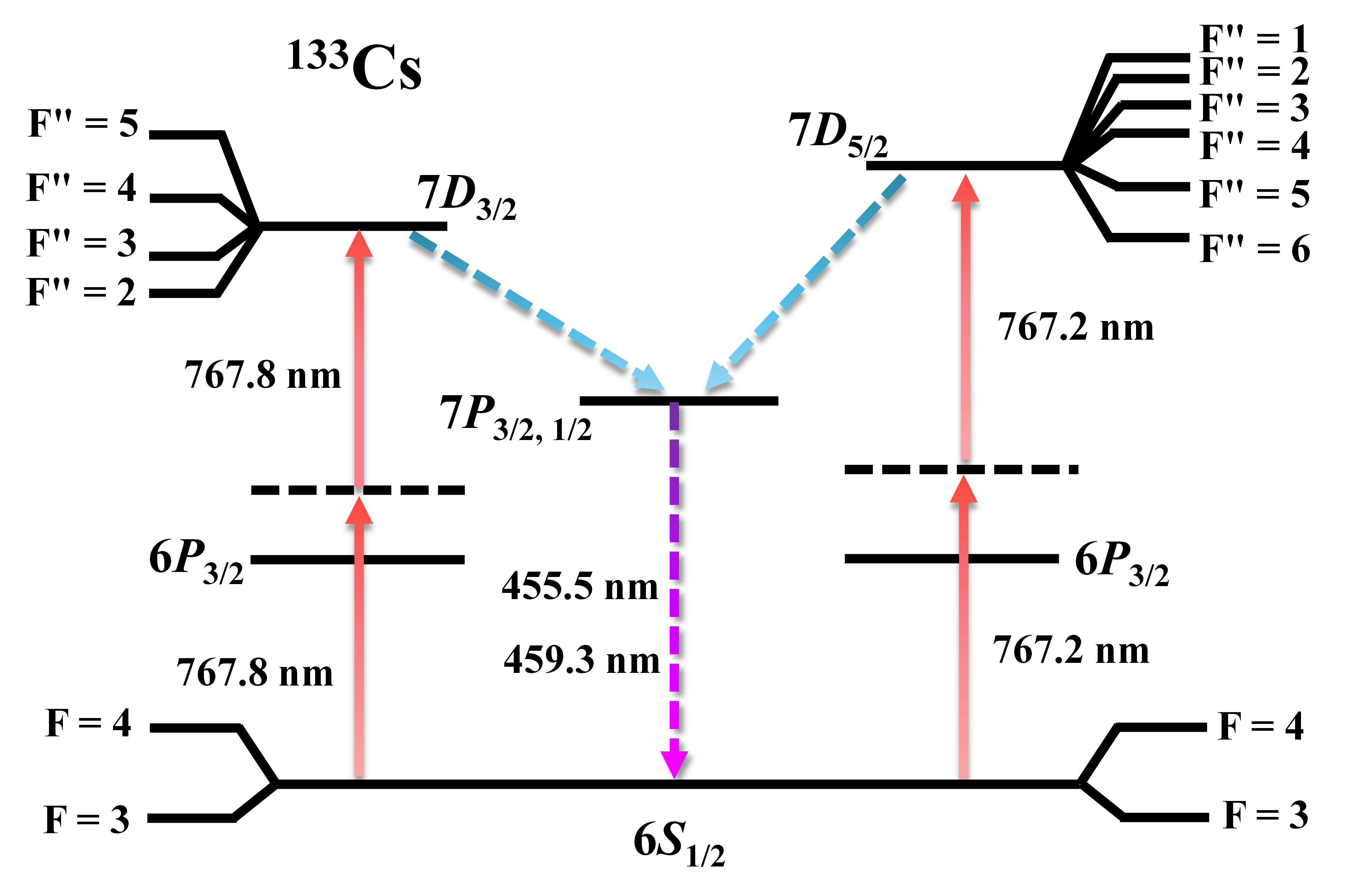}
 \centering
	\caption{Energy levels of $^{133}$Cs $6S_{1/2}-7D_{3/2, 5/2}$ monochromatic two-photon transitions. A 767 nm laser excites atoms from $6S_{1/2}$ state to $7D_{3/2, 5/2}$ states via a virtual level. The excited state atoms decay back to the $6S_{1/2}$ state through the $7P_{3/2, 1/2}$ intermediate states, then spontaneously emit 455.5 nm and 459.3 nm fluorescence (purple line).}
\end{figure}

\section{Experiment Setup}

The energy level diagram for the $6S_{1/2}-7D_{3/2, 5/2}$ of Cs is shown in Fig. 1. The atoms in a vapor are excited from the $6S_{1/2}$ ground state to the $7D$ excited states by absorbing two photons. Since the $^{133}$Cs nuclear spin $I$ is 7/2, there are two hyperfine structures $F=4$ and $F=3$ for the ground state $6S_{1/2}$ energy level. For the $6S-7D$ two-photon transition, the $6S_{1/2}$ ground state cannot be directly coupled to the $7D$ excited states because of the dipole-forbidden transition rules, and the transition selection rules must satisfy $\bigtriangleup$$F$=$\pm$2, $\pm$1, or 0. Therefore, there are four transition lines: $6S_{1/2} (F=4)-7D_{3/2} (F^{\prime\prime}= 2, 3, 4, 5)$, $6S_{1/2} (F=3)-7D_{3/2} (F^{\prime\prime}= 2, 3, 4, 5)$, $6S_{1/2} (F=4)-7D_{5/2} (F^{\prime\prime}= 6, 5, 4, 3, 2)$ and $6S_{1/2} (F=3)-7D_{5/2} (F^{\prime\prime}= 5, 4, 3, 2, 1)$. The transition process is as follows: Firstly, atoms are excited from $6S_{1/2}$ state via a virtual level (black dashed line) to the $7D_{3/2, 5/2}$ excited states by using 767.8 nm and 767.2 nm laser, respectively. Then, the unstable excited state atoms will decay back to the ground state via two pathways, the first is spontaneously radiate to the $7P_{3/2}$ state and then to the $6S_{1/2}$ ground state with 455.5 nm fluorescence, the second is spontaneously radiate to the $7P_{1/2}$ state and then to the $6S_{1/2}$ ground level with 459.3 nm fluorescence. The intensity of the fluorescence radiation is proportional to the population of $7D_{3/2, 5/2}$ excited state atoms, so the fluorescence intensity at 455.5 nm and 459.3 nm can characterize the intensity of the $6S_{1/2}-7D_{3/2, 5/2}$ two-photon transition.

The transition process is achieved by two laser beams with counter-propagation configuration acting on the Cs thermal vapor. The laser is provided by a Ti: sapphire laser system (SolaTis-SRX-XF, M Squared Lasers) which can be tuned from 600 to 1000 nm. The light emitted by the laser is linear polarized and has a lager laser power of 2 W. The laser frequency, which is locked by a reference cavity, can be scanned by changing the length of the reference cavity. A small fraction of the laser beam is fed to a wavelength meter in order to monitor the laser wavelength. A quartz tube of 100 mm length and 25 mm diameter is filled with cesium atoms, which is shielded with a $\mu$-metal box to reduce the effect of any stray magnetic field. The temperature of the vapor can be accurately controlled by a self-feedback system. An electro-optic modulator (EOM) generated a side-band laser beam is served as a frequency rule to measure hyperfine level intervals. The decay fluorescence at 455.5 nm and 459.3 nm are collected by a photomultiplier tube (PMT) (Hamamatsu, CR131), the high gain of $8\times10^{6}$ ensures the high efficiency spectral detection. An interference filter with center wavelength 457 nm and 10 nm pass band (FL457.9-10, Thorlabs) is placed in front of the PMT to prevent scattered light.

\section{Results and discussion}

The two-photon transition peak intensity from the ground state to the excited state $\mid$$n^{\prime\prime}$$L^{\prime\prime}_{J^{\prime\prime}}$$F^{\prime\prime}$$\rangle$ can be characterized by:\cite{Demtroder2008,Wang2015}

\begin{widetext}
\begin{equation}
\centering
\begin{split}
&P(6S_{1/2}F,n^{\prime\prime}L^{\prime\prime}_{J^{\prime\prime}}F^{\prime\prime})\propto(\frac{1}{2F+1})\frac{I_{1}I_{2}}{[\omega_{6S_{1/2}F:n^{\prime\prime}L^{\prime\prime}_{J^{\prime\prime}}F^{\prime\prime}}-(\omega_{1}+k_{1}\cdot v)-(\omega_{2}+k_{2}\cdot v)]^{2}+(\frac{\gamma_{n^{\prime\prime}L^{\prime\prime}_{J^{\prime\prime}}}}{2})^{2}}\\
&\times\Sigma_{M_{F},M_{F''}}|\Sigma_{F',M_{F'}}\frac{\langle n^{\prime\prime}L^{\prime\prime}_{J^{\prime\prime}}F^{\prime\prime}M''_{F}|\hat{e}_{2}\cdot d|6P_{J^{\prime\prime}}F'M'_{F}\rangle\langle6P_{J^{\prime\prime}}F'M'_{F}|\hat{e}_{1}\cdot d|6S_{1/2}FM_{F}\rangle}{\omega_{6S_{1/2}F:6P_{J^{\prime}}F'}-(\omega_{1}+k_{1}\cdot v)-i\frac{\gamma_{6P_{J^{\prime}}}}{2}}|^{2}
\end{split}
\end{equation}
\end{widetext}
where $I_{1}$ and $I_{2}$ are the intensities of two counter-propagating beams that used to excite the two-photon process, $F$, $F^{\prime}$ and $F^{\prime\prime}$ are total atomic angular momentum quantum numbers, $\omega_{1}$ and $\omega_{2}$ are the two photon frequencies, $k_{1}$ and $k_{2}$ are the wave vectors, $\upsilon$ is the atomic velocity, $\hat{e}_{1}$ and $\hat{e}_{2}$  are the unit vector along the quantization axis direction for the two laser beams, \emph{d} is the electric dipole operator, $\gamma_{nL}$is the homogeneous linewidth of the $|nL_{J}\rangle$ state, $M_{F}, M_{F}'$, and $M''_{F}$  are the magnetic quantum numbers. For a given laser polarization, the matrix elements for the different magnetic sublevels is related to reduced matrix elements by use of the Wigner-Eckart theorem and standard Clebsch-Gordan relations, which can be used to characterize the two-photon transition probability:\cite{Cheng2017,Sobelman1996}

\begin{equation}
\begin{split}
&\langle n'L'_{J'}F'M'_{F}| \hat{e} \cdot d|nL_{J}FM_{F}\rangle\\
&=\langle J||d||J'\rangle(-1)^{2F'+M_{{F}}+I+J}\sqrt{(2F+1)(2F'+1)(2J+1)}\\
&\times\left(\begin{array}{ccc}F'&1&F\\M'_{F}& M_{F}-M'_{F}&-M_{F}\end{array}\right)\left\{\begin{array}{ccc}J&J'&1\\F'&F&I\end{array}\right\}
\end{split}
\end{equation}%
where $\langle J\|d|J'\rangle$ is the reduced matrix element, $I$ is the nuclear spin, and the terms in brackets and curly brackets are the 3-$J$ and 6-$J$ symbols, respectively. For the $\pi$-$\pi$ polarization combination of Cs atoms, the hyperfine transition probabilities of the $6S_{1/2}(F=4)-7D_{3/2}(F^{\prime\prime}=2,3,4,5)$ and the $6S_{1/2}(F=3)-7D_{3/2}(F^{\prime\prime}=2,3,4,5)$ are 0.58 : 0.68 : 0.64 : 0.43 and 0.06 : 0.23 : 0.53 : 1, respectively. And, the hyperfine transition probabilities of the $6S_{1/2}(F=4)-7D_{5/2}(F^{\prime\prime}=6,5,4,3,2)$ and the $6S_{1/2}(F=3)-7D_{5/2}(F^{\prime\prime}=5,4,3,2,1)$ are 1 : 0.55 : 0.25 : 0.09 : 0.02 and 0.30 : 0.40 : 0.42 : 0.28 : 0.09, respectively.

\begin{figure}[ht!]
\centering
	\includegraphics[width=3.6in]{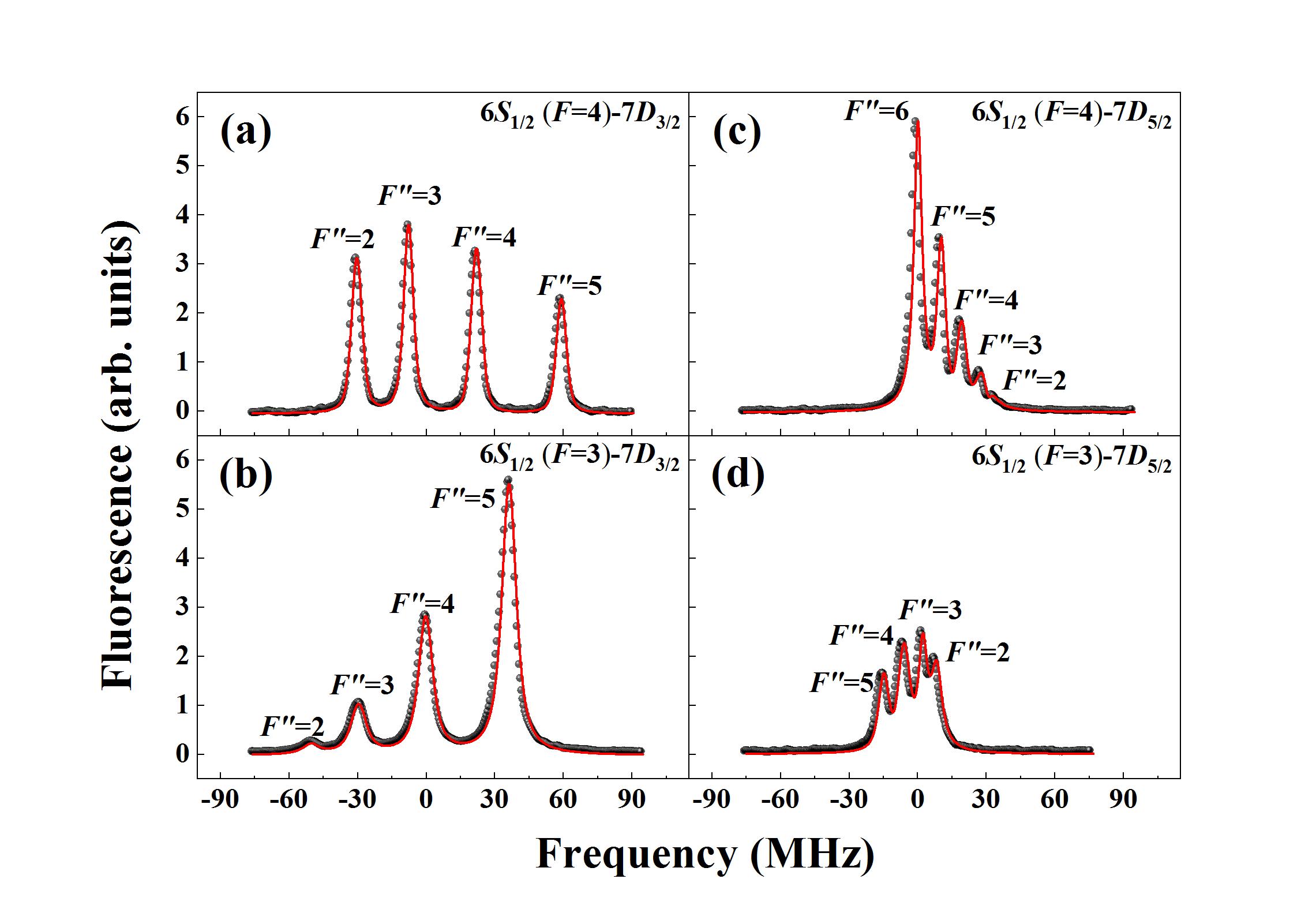}
	\caption{Monochromatic two-photon transition fluorescence spectrum (a) $6S_{1/2}(F=4)-7D_{3/2}(F^{\prime\prime}=2,3,4, 5)$, (b) $6S_{1/2}(F=3)-7D_{3/2}(F^{\prime\prime}=2,3,4,5)$, (c) $6S_{1/2}(F=4)-7D_{5/2}(F^{\prime\prime}=6,5,4,3,2)$, (d) $6S_{1/2}(F=3)-7D_{5/2}(F^{\prime\prime}=5,4,3,2)$. The black dots represent the experimental results. The red lines are multi-peak fitting curves with a Voigt function.}
\end{figure}

The two-photon transition fluorescence spectra of Cs $6S_{1/2}-7D_{3/2, 5/2}$ are obtained by scanning laser frequency, in which the two counter-propagation beams have same linear polarization. The experiment parameters are as follow: The laser power is 180 mW and the temperature of the atom vapor is fixed at 423 K. Figures 2(a) and (b) present the $6S_{1/2}(F=4)-7D_{3/2}(F^{\prime\prime}=2,3,4, 5)$ and $6S_{1/2}(F=3)-7D_{3/2}(F^{\prime\prime}=2,3,4,5)$ hyperfine transitions, respectively. Each hyperfine splitting of the $7D_{3/2}$ state can be well resolved. And the hyperfine intervals of $F^{\prime\prime}=2-F^{\prime\prime}=3$, $F^{\prime\prime}=3-F^{\prime\prime}=4$, $F^{\prime\prime}=4-F^{\prime\prime}=5$ are determined as 22.33 MHz, 29.70 MHz and 36.85 MHz, respectively. Figures 2(c) and (d) are the hyperfine transition spectra corresponding to $6S_{1/2}(F=4)-7D_{5/2}(F^{\prime\prime}=6,5,4,3,2)$ and $6S_{1/2}(F=3)-7D_{5/2}(F^{\prime\prime}=5,4,3,2)$, respectively. The $6S_{1/2}(F=3)-7D_{5/2}(F^{\prime\prime}=1)$ transition is not well resolved due to the small hyperfine separation with the neighboring one. And the hyperfine intervals of $F^{\prime\prime}=6-F^{\prime\prime}=5$, $F^{\prime\prime}=5-F^{\prime\prime}=4$, $F^{\prime\prime}=4-F^{\prime\prime}=3$ and $F^{\prime\prime}=3-F^{\prime\prime}=2$ are measured as 10.20 MHz, 9.01 MHz, 7.40 MHz and 5.82 MHz, respectively.

\begin{figure}[ht!]
	\includegraphics[width=3.5in]{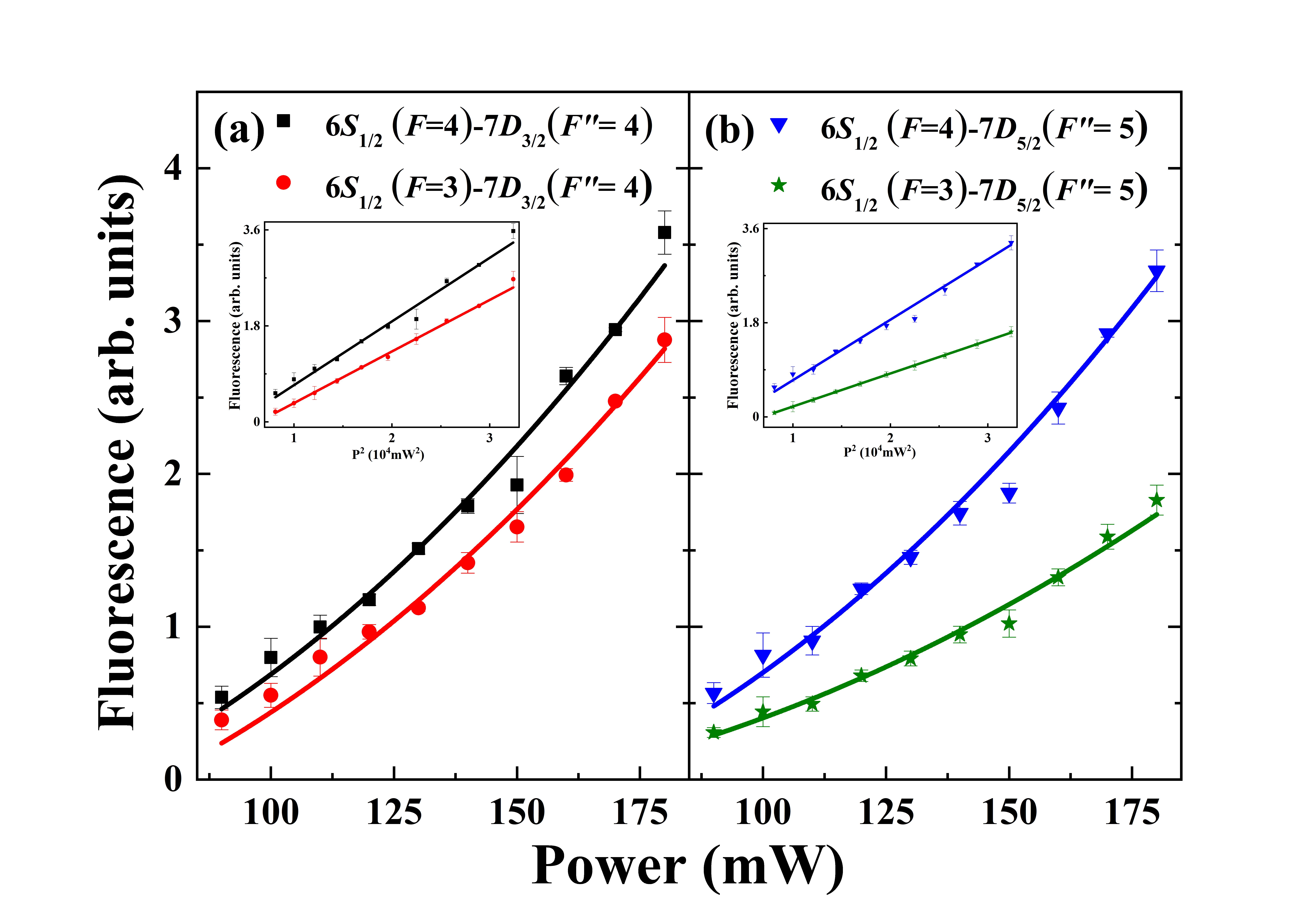}
 \centering
	\caption{The two-photon transitions fluorescence intensity as a function of laser power (a) $6S_{1/2}(F=4)-7D_{3/2} (F^{\prime\prime}=4)$ transition and $6S_{1/2}(F=3)-7D_{3/2} (F^{\prime\prime}=4)$ transition, (b) $6S_{1/2}(F=4)-7D_{5/2} (F^{\prime\prime}=5)$ transition and $6S_{1/2}(F=3)-7D_{5/2} (F^{\prime\prime}=5)$ transition. The dots represent experimental results, the solid lines are the results of theoretical fitting, and the errors are the standard deviation of three measurements. [Inset: the two-photon transition fluorescence intensity versus the squared total laser power ($P^{2}$)]. }
\end{figure}

 The two-photon transition fluorescence intensity is characterized by varying the laser power, which is shown in Fig. 3. We choose four different two-photon transitions for research: (a) $6S_{1/2}(F=4)-7D_{3/2} (F^{\prime\prime}=4)$ and $6S_{1/2}(F=3)-7D_{3/2} (F^{\prime\prime}=4)$ , and (b) $6S_{1/2}(F=4)-7D_{5/2} (F^{\prime\prime}=5)$  and $6S_{1/2}(F=3)-7D_{5/2} (F^{\prime\prime}=5)$. The experiment is conducted with the same condition of Fig. 2, except that the laser power increased from 90 mW to 180 mW. It can be seen that the two-photon transition intensity increases with the increase of laser power and exhibits a quadratic relationship. For the monochromatic two-photon transition with two counter-propagating laser beams, the laser intensity $I_{1}$ = $I_{2}$, the two-photon transition intensity has a linear relationship with the squared laser power from formula (1), which can be verified in the insert of Fig. 3. Meanwhile, the transition intensity ratio of $6S_{1/2}(F=4)-7D_{3/2} (F^{\prime\prime}=4)$ and $6S_{1/2}(F=3)-7D_{3/2} (F^{\prime\prime}=4)$ in Fig. 3(a) is 1.24, the ratio of $6S_{1/2}(F=4)-7D_{5/2} (F^{\prime\prime}=5)$  and $6S_{1/2}(F=3)-7D_{5/2} (F^{\prime\prime}=5)$in Fig. 3(b) is 1.84. These measured values match well with theoretical prediction of 1.22 and 1.83 by formula (2). Also, there is a small difference between experimental results and theoretical predictions, which is mainly coming from the systematic fitting errors in determining the two-photon transition intensity from the fluorescence spectroscopy and the slope between the fluorescence intensity and the squared laser power. Moreover, the spectral linewidth increases with increasing power due to the power-broadening effect\cite{He2013}.

\begin{figure}[ht!]
	\includegraphics[width=3.5in]{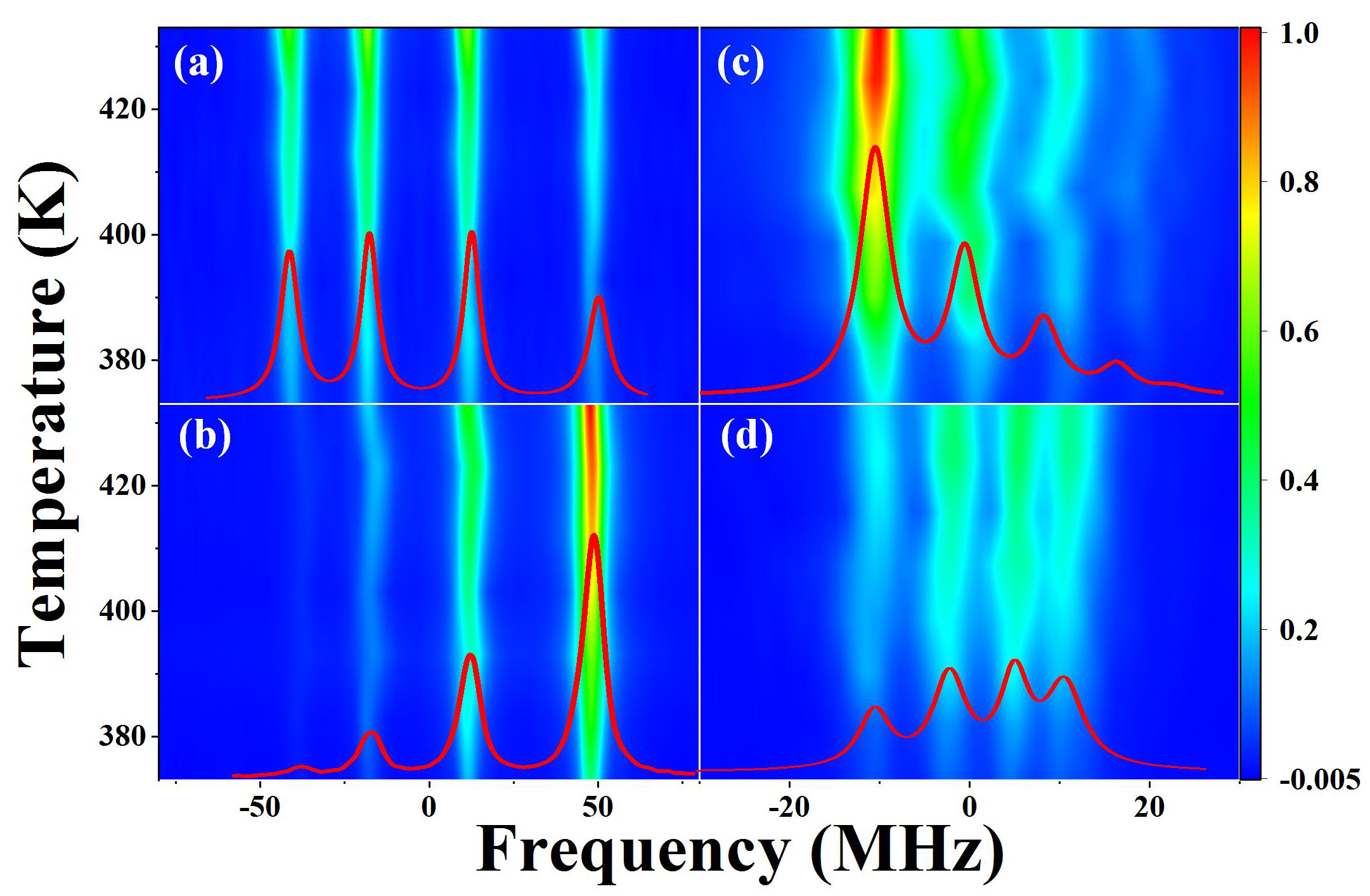}
	\caption{  The two-photon transition fluorescence intensity versus the vapor temperature. (a) $6S_{1/2}(F=4)-7D_{3/2}(F^{\prime\prime}=2,3,4, 5)$ transition , (b) $6S_{1/2}(F=3)-7D_{3/2}(F^{\prime\prime}=2,3,4, 5)$ transition, (c) $6S_{1/2}(F=4)-7D_{5/2}(F^{\prime\prime}=6,5,4,3,2)$ transition and (d) $6S_{1/2}(F=3)-7D_{5/2}(F^{\prime\prime}=5,4,3,2)$ transition. The red lines represent the two-photon transition spectra.}
\end{figure}

Figure 4 illustrates the two-photon transition intensity as a function of vapor temperature. The atom vapor temperature increased from 373 K to 433 K, and the corresponding atomic density is increased from $1.55\times10^{13}$  $cm^{-3}$ to 3.54$\times10^{14}$  $cm^{-3}$. The red lines are the transition spectra with the Fig. 2 for visual reference. Figures 4(a) and 4(b) are the experimental results of the $6S_{1/2}(F=4)-7D_{3/2}$ and $6S_{1/2}(F=3)-7D_{3/2}$ two-photon transitions cases, respectively. We can find that the fluorescence intensity of $6S_{1/2}(F=3)-7D_{3/2}$ is stronger than $6S_{1/2} (F=4)-7D_{3/2}$ transition at same temperature for the larger transition dipole matrix element. Figures 4(c) and 4(d) show the results of $6S_{1/2}(F=4)-7D_{5/2}$ and $6S_{1/2}(F=3)-7D_{5/2}$ transitions, respectively. Similarly, the intensity of $6S_{1/2}(F=4)-7D_{5/2}$  transition is obviously stronger than $6S_{1/2}(F=3)-7D_{5/2}$ transition at the same temperature just as the theory. The increase of the fluorescence intensity with the vapor temperature attributed to the increasing atomic density. However, the fluorescence intensity remains constant when the temperature is higher than 423 K for the self-absorption of the $7P_{1/2, 3/2}-6S_{1/2}$ transition \cite{Ryan1993}. Also, the linewidth of the transition spectrum is found to be a synchronous increase with the fluorescence intensity as the vapor temperature increases due to the Doppler broadening effect \cite{Wang2019}.

The laser polarization combinations have a direct influence on the two-photon transition probability for the different transition pathways. The different magnetic sublevel transitions of $\triangle$$m_{F}$= 0, 1 and -1 are driven by $\pi$ linearly polarized, $\sigma^{+}$ right-handed circularly polarized and $\sigma^{-}$ left-handed circularly polarized laser, respectively. The $6S-7D$ two-photon transition probabilities for each allowed magnetic sublevel levels can be obtained by multiplying probabilities for the two magnetic sublevel transitions. In the experiment setup, the insert of different quarter-waveplate before or behind the atom vapor can result in different polarization combinations of laser beams. Figure 5 shows the experimental results, where the black, red and blue lines represent the cases of $\sigma^{+}$-$\sigma^{+}$, $\pi$-$\pi$, and $\sigma^{+}$-$\sigma^{-}$ polarization combinations, respectively. We find that the $\sigma^{+}$-$\sigma^{+}$ polarization combination induced the biggest transition intensity, the $\pi$-$\pi$ case is smaller, while the opposite circular polarization combination causes the minimum transition probability. These results are consistent with the theoretical prediction, the relative probabilities of two-photon transition tend to be maximum and minimum in the case of the same rotation circular polarization and the opposite rotation circular polarization, respectively\cite{Dai2016,McGloin2000}.

\begin{figure}[ht!]
	\includegraphics[width=3.5in]{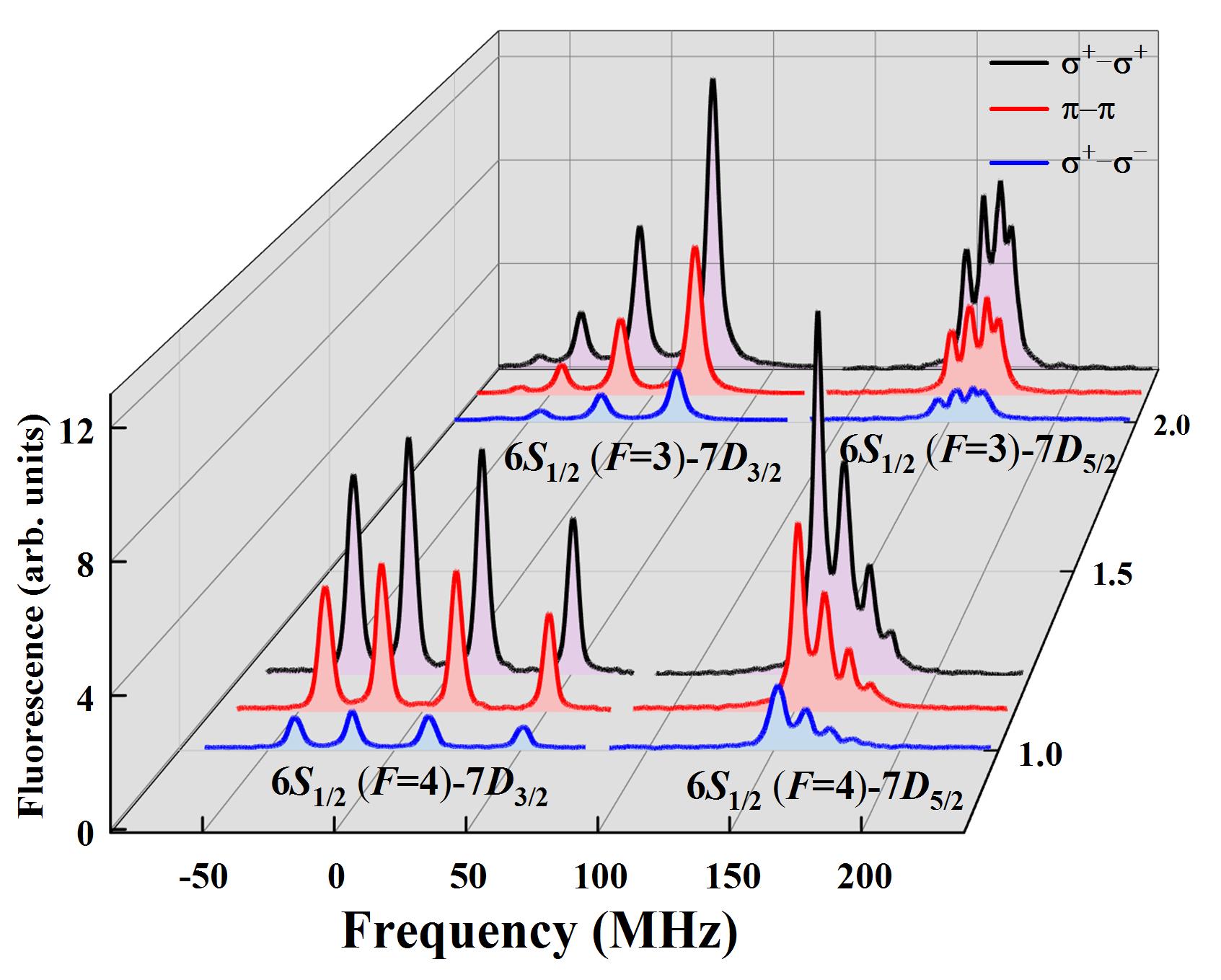}
 \centering
	\caption{ The $6S_{1/2} -7D_{3/2, 5/2}$ two-photon transition spectra with different polarization combinations of the laser beams. The black, red and blue lines are the experimental results of $\sigma^{+}$-$\sigma^{+}$, $\pi$-$\pi$, and $\sigma^{+}$-$\sigma^{-}$ polarization combinations, respectively.}
\end{figure}

\begin{table*}
 \centering
    \caption{Hyperfine coupling constants $A$ and $B$ of the $7D_{3/2}$ and $7D_{5/2}$ states}
\setlength{\tabcolsep}{6mm}
\begin{tabular}{cccl} % 控制表格的格式
  \hline
  \hline
  Hyperfine state & $A$ & $B$ & Reference\\
  \hline
              &7.39(0.06)&-0.19(0.18)&This work\\
              &7.36(0.03)&-0.10(0.20)&Kortyna \emph{et al.}(2008)\cite{Kortyna2008}\\
  {$7D_{3/2}$}&7.386(0.015)&-0.18(0.16)&Stalnaker \emph{et al.}(2010)\cite{Stalnaker2010}\\
              &7.36(0.07)&-0.88(0.87)&Lee \emph{et al.}(2011)\cite{Lee2011}\\
              &7.38(0.01)&-0.18(0.10)&Kumar \emph{et al.}(2013)\cite{Kumar2013}\\
  \hline
              &-1.79(0.05)&1.05(0.29)&This work\\
  {$7D_{5/2}$}&-1.42&-&Auzinsh \emph{et al.}(2007)\cite{Auzinsh2007}\\
              &-1.717(0.015)&-0.18(0.52)&Stalnaker \emph{et al.}(2010)\cite{Stalnaker2010}\\
              &-1.81(0.05)&1.01(1.06)&Lee \emph{et al.}(2011)\cite{Lee2011}\\
  \hline
  \hline
  \end{tabular}
\end{table*}

The hyperfine interaction can be written in terms of the hyperfine energy shift \cite{Wang2014}:

\begin{equation}
\triangle E_{hfs}=\frac {A \cdot K}{2} + \frac{B}{2}\cdot\frac{\frac{3K}{2}(K+1)-2I(I+1)J(J+1)}{I(2I-1)J(2J-1)}.
\end{equation}%
where $A$ is the magnetic dipole constant, $B$ is the electric quadruple constant, $I$ is the nuclear spin angular momentum quantum number, and $J$ is the total electron angular momentum quantum number. The total atomic angular moment is $F = I + J$, and $K=F(F+1)-J(J+1)-I(I+1)$. With the hyperfine structure splittings measured from the Figs. 2 (a-d), the hyperfine coupling constants $A$ and $B$ of Cs atoms $7D_{3/2, 5/2}$ states are determined by using formula (3), which are listed in TABLE I. Our experimental results are in good agreement with previous reports and are more consistent with the measurement results by direct frequency comb spectroscopy \cite{Stalnaker2010}. However, compared with direct frequency comb spectroscopy, the monochromatic two-photon transition is expected to benefit for the better elimination of the Doppler effect, higher spectral SNR, and convenient system integration and miniaturization.

In the whole measurement process, the error source mainly comes from systematic uncertainties, such as Zeeman shift, AC-stark shift, pressure shift, frequency drift of laser. In our experiment, a $\mu$-metal box is used to minimize the stray magnetic field by surrounding the Cs vapor. Thus, the Zeeman shift can almost be eliminated. The AC-stark displacement is the frequency shift of the atomic energy level due to the dipole action of the light field on the atom-induced electric dipole moment. The energy level frequency shift will cause measurement error, which is proportional to the laser intensity. For the operating laser power of our experiment, the AC-Stark shift is calculated to be about 83 kHz. The pressure shift of the system is caused by the Cs atom vapor. From the detailed study of pressure shifts in Ref. \cite{Stalnaker2010}, we expect the pressure shift of our system to be $\sim$140 kHz. The error caused by the frequency drift of the laser can be eliminated by multiple measurements. Another possible systematic error comes from the accuracy in determining the hyperfine structure by fitting the two-photon transition spectrum with the theoretical model, which is about ~30 kHz. All in all, the entire experimental system caused $\sim$250 kHz systematic uncertainty for the hyperfine structure measurement.\\

\section{Conclusion}

In conclusion, we have experimentally demonstrated the monochromatic $6S_{1/2} -7D_{3/2, 5/2}$ electric quadrupole transition in a thermal Cs vapor. A quadratic relationship between the laser power and the two-photon transition intensity is observed. Also, the fluorescence intensity shows an upward trend as the vapor temperature increases. The effect of polarization combinations of the counter-propagating laser beams on the transition spectrum is studied in detail. The $7D_{3/2, 5/2}$ states hyperfine structure constants $A$ and $B$ are derived by using the measured hyperfine level splitting intervals. The simple optical setup is easy to miniaturize and can be readily integrated into more complex devices. Moreover, the two-photon transition of Cs atoms at 767 nm can be used as a secondary frequency standard in the C-band window of quantum telecommunication.

\section{Acknowledgment}

This work is supported by the National Key R$\&$D Program of China under Grant No. 2017YFA0304203; the NSFC under Grants No. 61875112, No. 61705122, No. 91736209; the Program for Sanjin Scholars of Shanxi Province; Key Research and Development Program of Shanxi Province for International Cooperation under Grant No. 201803D421034 and 1331KSC.


\begin{thebibliography}{99}

\bibitem{Wood1997}
C. S. Wood, S. C. Bennett, D. Cho, B. P. Masterson, J. L. Roberts, C. E. Tanner, and C. E. Wieman, Measurement of parity non-conservation and an anapole moment in cesium, Science \textbf{275}, 1759 (1997).

\bibitem{Porsev2009}
S. G. Porsev, K. Beloy, and A. Derevianko, Precision determination of electroweak coupling from atomic parity violation and implications for particle physics, Phys. Rev. Lett. \textbf{102}, 181601 (2009).

\bibitem{Zheng2017}
X. Zheng, Y. Sun, J. Chen, W. Jiang, K. Pachucki, and S. Hu, Measurement of the frequency of the $2^{3}S - 2^{3}P$ transition of $^{4}$He, Phys. Rev. Lett. \textbf{119}, 263002 (2017).

\bibitem{Yuan2019}
J. Yuan, C. Wu, Y. Li, L. Wang, Y. Zhang, L. Xiao, S. Jia, Controllable electromagnetically induced grating in a cascade-type atomic system, Front. Phys. \textbf{14}, 52603 (2019).

\bibitem{Li2020}
R. Li, Y. Wu, Y. Rui, B. Li, Y. Jiang, L. Ma, and H. Wu, Absolute frequency measurement of ${^6}$Li $D$ lines with khz-Level uncertainty, Phy. Rev. Lett. \textbf{124}, 063002 (2020).

\bibitem{Ramos2019}
A. Ramos, R. Cardman, and G. Raithel, Measurement of the hyperfine coupling constant for $nS_{1/2}$ Rydberg states of $^{85}$Rb, Phys. Rev. A \textbf{100}, 062515 (2019).

\bibitem{Arimondo1977}
E. Arimondo, M. Inguscio, and P. Violino, Experimental determinations of the hyperfine structure in the alkali atoltls, Rev. Mod. Phys. \textbf{49}, 31 (1977).

\bibitem{Udem2000}
J. Yuan, C. Wu, L. Wang, G. Chen, and S. Jia, Observation of diffraction pattern in two-dimensional optically induced atomic lattice, Opt. Lett. \textbf{44}, 4123 (2019).

\bibitem{Kirkbride2014}
J. Kirkbride, A. R. Dalton, and G. A. D. Ritchie, Polarization spectroscopy of a velocity-selected molecular sample, Opt. Lett. \textbf{39}, 2645 (2014).

\bibitem{J. Yuan2019}
J. Yuan, S. Dong, C. Wu, L. Wang, L. Xiao, and S. Jia, Optically tunable grating in a $V + \Xi$ configuration involving a Rydberg state, Opt. Express \textbf{28}, 23820 (2020).

\bibitem{Ray2020}
T. Ray, R. K. Gupta1, V. Gokhroo1, J. L. Everett, T. Nieddu, K. S Rajasree and S. N. Chormaic, Observation of the $^{87}$Rb $5S_{1/2}$ to $4D_{3/2}$ electric quadrupole transition at 516.6 nm mediated via an optical nanofibre, New J. Phys. \textbf{22}, 062001 (2020).

\bibitem{Safronova2016}
M. S. Safronova, U. I. Safronova, and C. W. Clark, Magic wavelengths, matrix elements, polarizabilities, and lifetimes of Cs, Phy. Rev. A \textbf{94}, 012505 (2016).

\bibitem{Heshamia2016}
K. Heshamia, D. G. Englanda, P. C. Humphreysb, P. J. Bustarda, V. M. Acostac, J. Nunnb, and B. J. Sussmana, Quantum memories: emerging applications and recent advances, J Mod Opt. \textbf{63}, 2005 (2016).

\bibitem{Auzinsh2006}
M. Auzinsh, K. Blushs, R. Ferber, F. Gahbauer, A. Jarmola, and M. Tamanis, Electric field induced hyperfine level-crossings in (nD) Cs at two-step laser excitation: Experiment and theory, Opt. Commun. \textbf{264}, 333 (2006).

\bibitem{Auzinsh2007}
M. Auzinsh, K. Bluss, R. Ferber, F. Gahbauer, A. Jarmola, M. S. Safronova, U. I. Safronova, and M. Tamanis, Level crossing spectroscopy of the 7, 9, and 10$D_{5/2}$ states of $^{133}$Cs and validation of relativistic many-body calculations of the polarizabilities and hyperfine constants, Phys. Rev. A \textbf{75}, 022502 (2007).

\bibitem{Kortyna2008}
A. Kortyna, V. Fiore, and J. Farrar, Measurement of the cesium $7d^{2}D_{3/2}$ hyperfine coupling constants in a thermal beam using two-photon fluorescence spectroscopy, Phys. Rev. A \textbf{77}, 062505 (2008).

\bibitem{Stalnaker2010}
J. E. Stalnaker, V. Mbele, V. Gerginov, T. M. Fortier, S. A. Diddams, L. Hollberg, and C. E. Tanner, Femtosecond frequency comb measurement of absolute frequencies and hyperfine coupling constants in cesium vapor, Phys. Rev. A \textbf{81}, 043840 (2010).

\bibitem{Lee2011}
Y. Lee, Y. Chang, Y. Chang, Y. Chen, C. Tsai, and H. C. Chui, Hyperfine coupling constants of cesium $7D$ states using two-photon spectroscopy, Appl. Phys. B \textbf{105}, 391 (2011).

\bibitem{Kumar2013}
P. V. Kiran Kumar, M. Sankari, and M. V. Suryanarayana, Hyperfine structure of the $7d^{2}D_{3/2}$ level in cesium measured by Doppler-free two-photon spectroscopy, Phys. Rev. A \textbf{87}, 012503 (2013).

\bibitem{Demtroder2008}
W. Demtr\"{o}der, Laser Spectroscopy Vol. 2: Experimental Techniques 4th edn (Springer, Berlin, 2008).

\bibitem{Wang2015}
L. Wang, Y. Zhang, S. Xiang, S. Cao, L. Xiao, and S. Jia, Two-photon spectrum of $^{87}$Rb using optical frequency comb, Chin. Phys. B \textbf{24}, 063201(2015).

\bibitem{Cheng2017}
H. Cheng, H. Wang, S. Zhang, P. Xin, J. Luo, and H. Liu, Electromagnetically induced transparency of $^{87}$Rb in a buffer gas cell with magnetic field, J. Phys. B \textbf{50}, 095401 (2017).

\bibitem{Sobelman1996}
I. I. Sobelman, Atomic Spectra and Radiative Transitions (Springer,Berlin, 1996).

\bibitem{He2013}
Z. He, J. Tsai, Y. Chang, C. Liao, and C. Tsai, Ladder-type electromagnetically induced transparency with optical pumping effect, Phys. Rev. A \textbf{87}, 033402 (2013).

\bibitem{Ryan1993}
R. E. Ryan, L. A. Westling, and H. J. Metcalf, Two-photon spectroscopy in rubidium with a diode laser, J. Opt. Soc. Am. B \textbf{10}, 1643 (1993).

\bibitem{Wang2019}
S. Wang, J. Yuan, L. Wang, L. Xiao, and S. Jia, A stable frequency standard based on the one-color two-photon $5S-7S$ transition of rubidium at 760 nm, Laser Phys. Lett. \textbf{16} 125204 (2019).

\bibitem{Dai2016}
S. Dai, W. Xia, Y. Zhang, J. Zhao, D. Zhou, Q. Wang, Q. Yu, K. Li, X. Qi, and X. Chen, Polarization dependence of the direct two photon transitions of $^{87}$Rb atoms by erbium: Fiber laser frequency comb, Opt. Commun. \textbf{378}, 35 (2016).

\bibitem{McGloin2000}
D. McGloin, M. H. Dunn, and D. J. Fulton, Polarization effects in electromagnetically induced transparency, Phys. Rev. A \textbf{62}, 053802 (2000).

\bibitem{Wang2014}
J. Wang, H. Liu, G. Yang, B. Yang, and J. Wang, Determination of the hyperfine structure constants of the $^{87}$Rb and $^{85}$Rb $4D_{5/2}$ state and the isotope hyperfine anomaly, Phys. Rev. A \textbf{90}, 052505 (2014).















\end{thebibliography}
\end{document}